\documentclass[]{spie}  

\usepackage{amsmath,amsfonts,amssymb}
\usepackage{graphicx}
\usepackage[colorlinks=true, allcolors=blue]{hyperref}
\usepackage{enumitem}
\usepackage{multirow}
\usepackage{subcaption}
\usepackage{cleveref}
\usepackage{makecell}
\usepackage{bm}

\setlist[itemize]{noitemsep, topsep=0pt}
\setlist[enumerate]{noitemsep, topsep=0pt}

\title{Autonomous on-board data processing and instrument calibration software for the SO/PHI}

\author[a]{K. Albert}
\author[a]{J. Hirzberger}
\author[a]{D. Busse}
\author[b]{T. Lange}
\author[a]{M. Kolleck}
\author[b]{B. Fiethe}
\author[c]{D. Orozco Su\'{a}rez}
\author[a]{J. Woch}
\author[a]{J. Schou}
\author[d]{J. Blanco Rodr\'{i}guez}
\author[a]{A. Gandorfer}
\author[b]{Y. Guan}
\author[c]{J.P. Cobos Carrascosa}
\author[d]{D. Hern\'{a}ndez Exp\'{o}sito}
\author[c]{J.C. del Toro Iniesta}
\author[a]{S. K. Solanki}
\author[b]{H. Michalik}

\affil[a]{Max Planck Institute for Solar System Research, Justus-von-Liebig-Weg 3, G\"{o}ttingen, Germany}
\affil[b]{Institute of Computer and Network Engineering at the TU Braunschweig, Hans-Sommer-Stra$\ss$e 66, Braunschweig, Germany}
\affil[c]{Instituto de Astrof\'{i}sica de Andaluc\'{i}a (IAA - CSIC), Apartado 3004, Granada, Spain}
\affil[d]{Universidad de Valencia, Catedr\'{a}tico Jos\'{e} Beltr\'{a}n 2, Paterna (Valencia), Spain}

\authorinfo{Further author information: (Send correspondence to K.A.)\\K.A.: albert@mps.mpg.de, +49 551 384 979-186\\  J.H.: hirzberger@mps.mpg.de\\ D.B.: busse@mps.mpg.de}

\begin{document} 
\maketitle

\begin{abstract}
The extension of on-board data processing capabilities is an attractive option to reduce telemetry for scientific instruments on deep space missions. The challenges that this presents, however, require a comprehensive software system, which operates on the limited resources a data processing unit in space allows.\\
We implemented such a system for the Polarimetric and Helioseismic Imager (PHI) on-board the Solar Orbiter (SO) spacecraft. It ensures autonomous operation to handle long command-response times, easy changing of the processes after new lessons have been learned and meticulous book-keeping of all operations to ensure scientific accuracy. This contribution presents the requirements and main aspects of the software implementation, followed by an example of a task implemented in the software frame, and results from running it on SO/PHI.\\
The presented example shows that the different parts of the software framework work well together, and that the system processes data as we expect. The flexibility of the framework makes it possible to use it as a baseline for future applications with similar needs and limitations as SO/PHI.
\end{abstract}

\section{Introduction}
\begin{figure}[t]
	\centering
	\includegraphics[width=.6\textwidth]{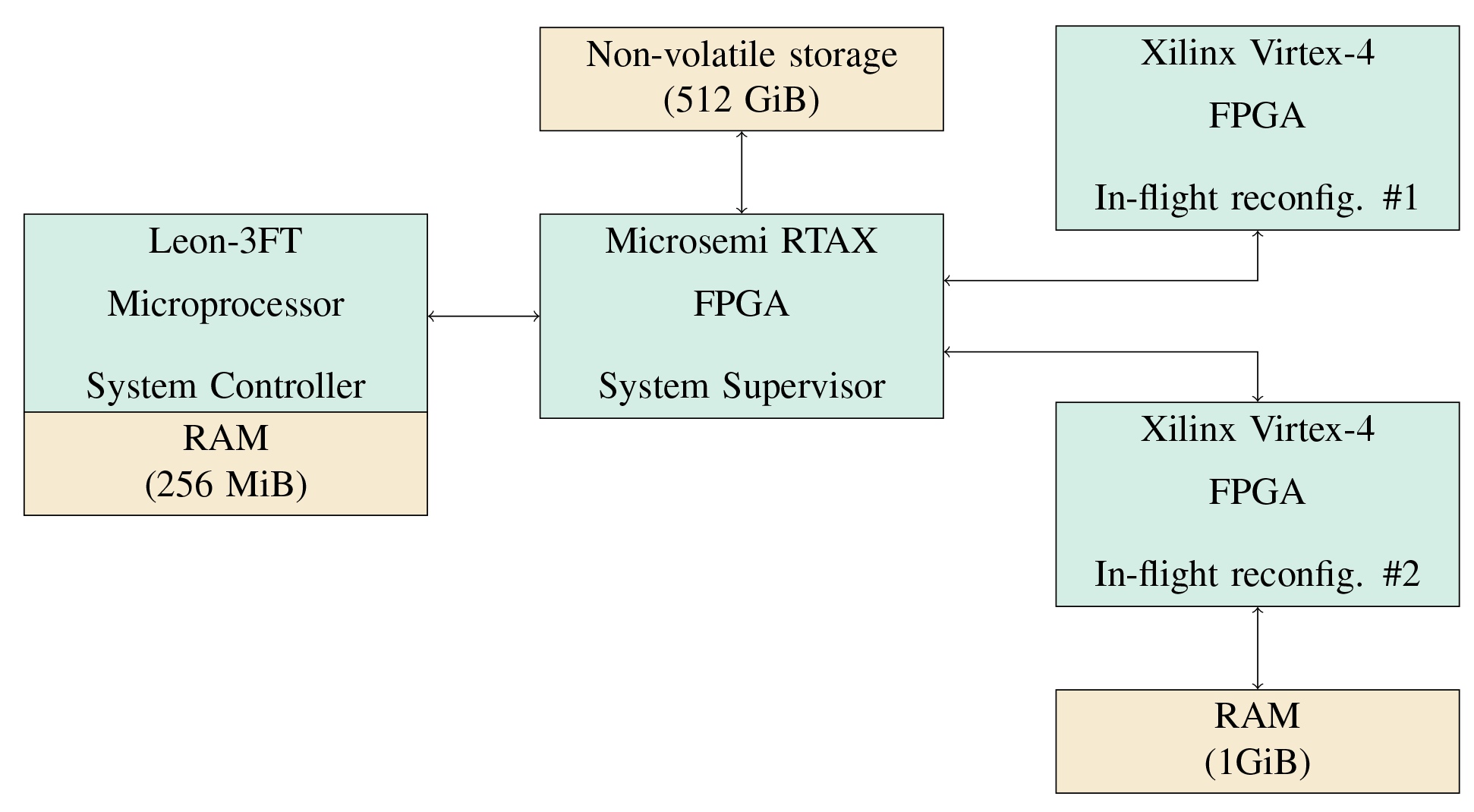}
	\vspace{.2 cm}
	\caption{The data processing of SO/PHI is distributed between a Leon-3FT microprocessor and Xilinx Virtex-4 RFPGAs. These units are aided by memories of different capacities. Data is stored in the non-volatile memory.}
	\label{fig:HW}
\end{figure}

Many state of the art scientific space instruments produce more data than what is possible to download to ground. This discrepancy is especially significant when, due to the orbit design, the mission can only ensure a low amount of telemetry. One way to reduce the necessary telemetry is to extend the on-board processing capabilities of the instrument.\\
This strategy is adapted for the Polarimetric and Helioseismic Imager (PHI)\cite{gandorfer2011solar}. PHI is the first imaging solar spectropolarimeter on-board a deep space mission: the Solar Orbiter (SO). It will image the solar photosphere in the light of the Fe I 617.3\,nm absorption line, at six wavelengths and in four polarisation states of light. This spectral line is sensitive to the magnetic field (through the Zeeman effect) and to the line of sight (LOS) velocities in the photosphere (due to the Doppler effect). Through the measurements described we can determine characteristics of the magnetic field and the LOS velocity at the average of the formation height for the absorption line. To retrieve these quantities the polarised radiative transfer equation (RTE) must be inverted\cite{iniesta2016inv}.\\
SO/PHI implements on-board autonomous instrument calibration and on-board autonomous science data analysis including RTE inversion. This means that the calibration data can be directly applied on-board to the science observables without adding to the telemetry volume. Therefore the necessary telemetry is reduced to the continuum intensity, the magnetic field vector and the LOS velocity instead of the raw science observables. These functions are implemented for the first time on this type of instrument with the limited computational resources a space instrument offers. Comparable systems have been implemented on the AMPTE IRM 3D Plasma Instrument, the Giotto RPA Experiment\cite{curtis1989board} and the SOHO/MDI\cite{scherrer1995solar}. However, all these cases are considerably more limited in capability than the system implemented for SO/PHI.\\
The hardware available for data processing\cite{fiethe2007reconfigurable}$^,$\cite{fiethe2012adaptive} integrates a system controller microprocessor, a system supervisor Field Programmable Gate Array (FPGA), and two dynamically reconfigurable FPGAs (RFPGAs). See \cref{fig:HW}. The processing is aided with a 1\,GiB processing memory connected to one of the RFPGAs, and a much smaller, 256\,MiB memory connected to the microprocessor, of which only a part is available for image processing. A non-volatile storage of 512\,GiB is available for storing data in the instrument. The in flight reconfigurability of the two Xilinx FPGAs enables us to take a time-sharing approach, saving volume, mass and energy.\\
SO/PHI captures science images during dedicated observation windows along the orbits, each lasting for 10\,days. An in-orbit calibration campaign is associated with each observation window with the aim of characterising the instrument in the observational conditions. The data processing is done after the observation window, using the data collected in the calibration campaign. The planning of the operations and the data processing is done together and on long term, well ahead of the time of their execution. This includes the specification of all input and output addresses, as well as the selection of the correct calibration data for the reduction of the dataset.\\
In the following we describe the software frame for the on-board data calibration and on-board data pre-processing steps performed before the inversion of the RTE.

\section{Requirements of the data processing system}
The data processing system of SO/PHI performs three different tasks. It: 
\begin{itemize}
	\item determines the optimal operational parameters for the instrument (e.g. focus position) from dedicated datasets, refining the result in a second observation,
	\item calculates the calibration data from datasets obtained during dedicated observations,
	\item processes the raw science observables by removing the instrumental effects, inverting the RTE and compressing the data.
\end{itemize}
The processing tasks involve computationally demanding image processing steps. To shorten run time, these functions are implemented in one of the two RFPGAs. However, a back-up solution, implemented in the microprocessor is also required to facilitate early testing during development, as well as to contribute to fault tolerance. This results in the need of ensuring that the same task can run on two platforms with different amount of processing memory.\\
To reduce the implementation complexity of the RFPGA functions, a fixed point notation has been adapted for the processing system. To maintain the requirements for scientific accuracy the images are rescaled before the processing steps are applied to them.\\
Due to the novelty of the mission a number of questions related to the data processing steps may only be answered after first light. This requires the possibility for easy changes in the task implementations.\\
Due to the autonomous execution and no access to the intermediate results, the only record of the operations is kept in a processing log file, as part of the metadata of each dataset. The metadata recording therefore must be integral and secure, as well as provide data that facilitates error search and improvement of the algorithms.

\section{Implementation of the data processing system}

We take a unified approach to the tasks, implementing them in the form of pipelines. A pipeline is defined to be a series of operations executed on the same target (a dataset or a subset of a dataset). All pipelines start with loading the target from the non-volatile memory into the processing memory, and end by either storing the result dataset in the non-volatile memory, or by returning a calculated parameter.\\
To implement the computationally demanding functions on the RFPGAs we define several FPGA configurations, each of them containing dedicated hardware logic that implements parts of the on-board processing (e.g. RTE inversion, or a group of different image processing functions), and load them into the RFPGAs as needed.\cite{lange2017board}\\
To fulfil the requirement on easily changing the pipelines a block approach is taken. The blocks access the image processing functions running on the RFPGAs, or the microprocessor in back-up implementation, and combine them into useful steps. To facilitate changing their order, they have a unified interface. Within the pipeline all blocks have the same target, loaded as the first step of the pipeline into the processing memory. The blocks ensure that the target at their interface is scaled to be represented on the maximum number of available bits. All blocks may load additional data necessary for the processing steps, which are discarded at their termination. Each block must also determine the dataset history, to facilitate their free ordering. The parameters necessary for this are not passed as variables in-between the blocks, instead they are transferred through the metadata. All blocks are executed sequentially, and a pipeline is manipulating the target only through blocks.\\
Due to the requirement on executing the pipeline both with and without the RFPGAs, the pipeline must run with different amounts of processing memory. In the back-up solution, the memory cannot hold a full dataset. Therefore, the pipeline is parametrised and can be configured to run on a subset of the full dataset. In many steps the pipeline needs one image of the dataset. However, there are exceptions in which corresponding pixels from multiple images are necessary. Consequently, we split the pipeline into an image linear and image parallel part, each running on a part of the full dataset: a number of images or a number of rows from all images.\\
Error handling is done with two error levels: errors and warnings, based on their severity. Errors interrupt the processing, while warnings are checked on ground to evaluate the correctness of the results. Both are marked in the return value of all operations, blocks and pipelines. In order to log multiple warnings and errors detected, the return values are defined as bit masks.\\
The metadata of a dataset contains all relevant information and is recorded at different times: at image acquisition, and at data processing. The information is organised by entries, containing parameters that form a logical unit (e.g. the return value and parameters of the image processing functions), and the metadata file is treated as a log. 

\section{Example pipeline and its run}
\begin{figure}[p]
	\centering
	
	\begin{subfigure}[b]{.47\textwidth}
		
		\includegraphics[width=\textwidth]{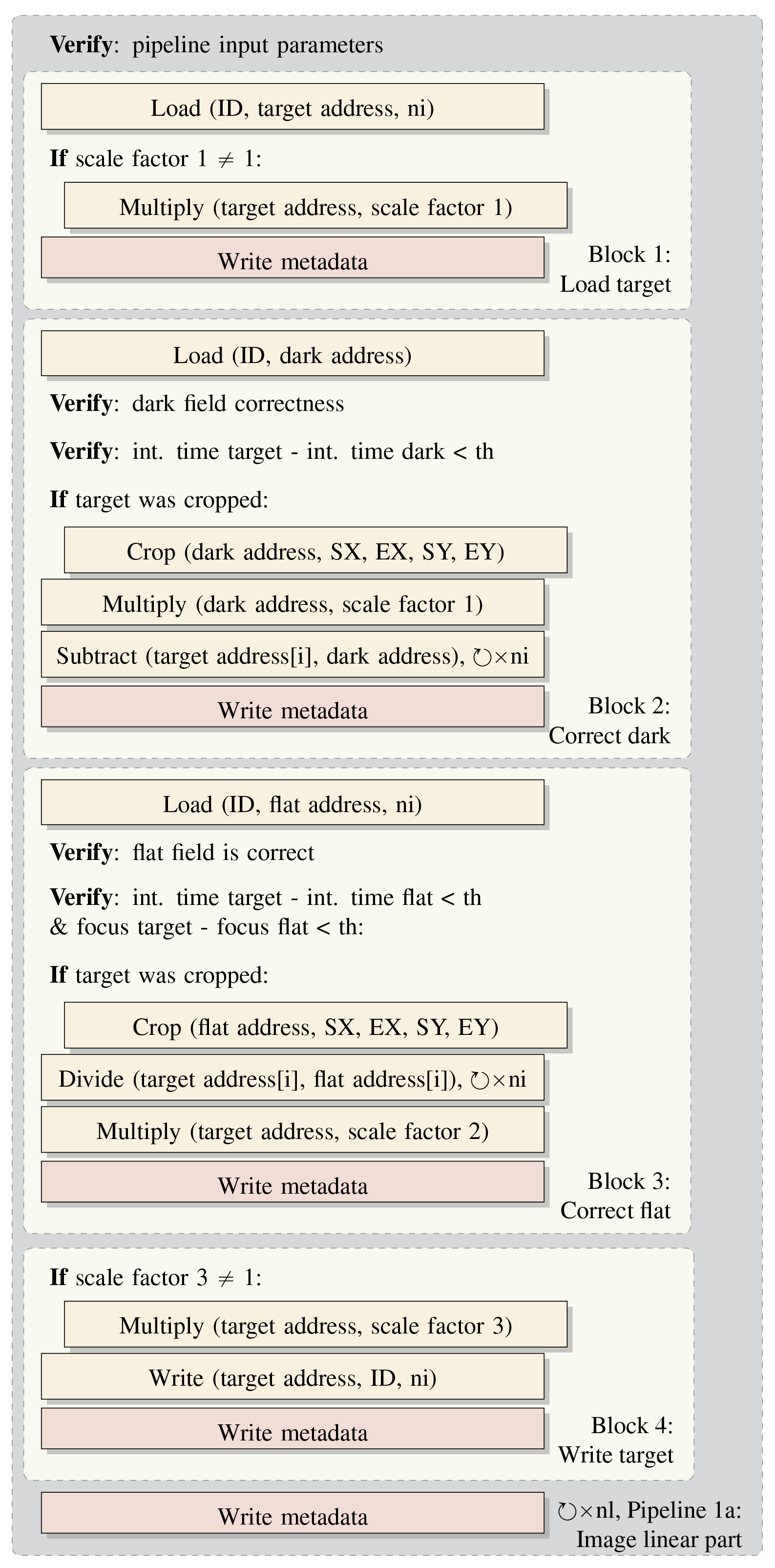}
		\caption{Image linear part.}
		\label{fig:lin}
	\end{subfigure}
	\begin{subfigure}[b]{.47\textwidth}
		
		\includegraphics[width=\textwidth]{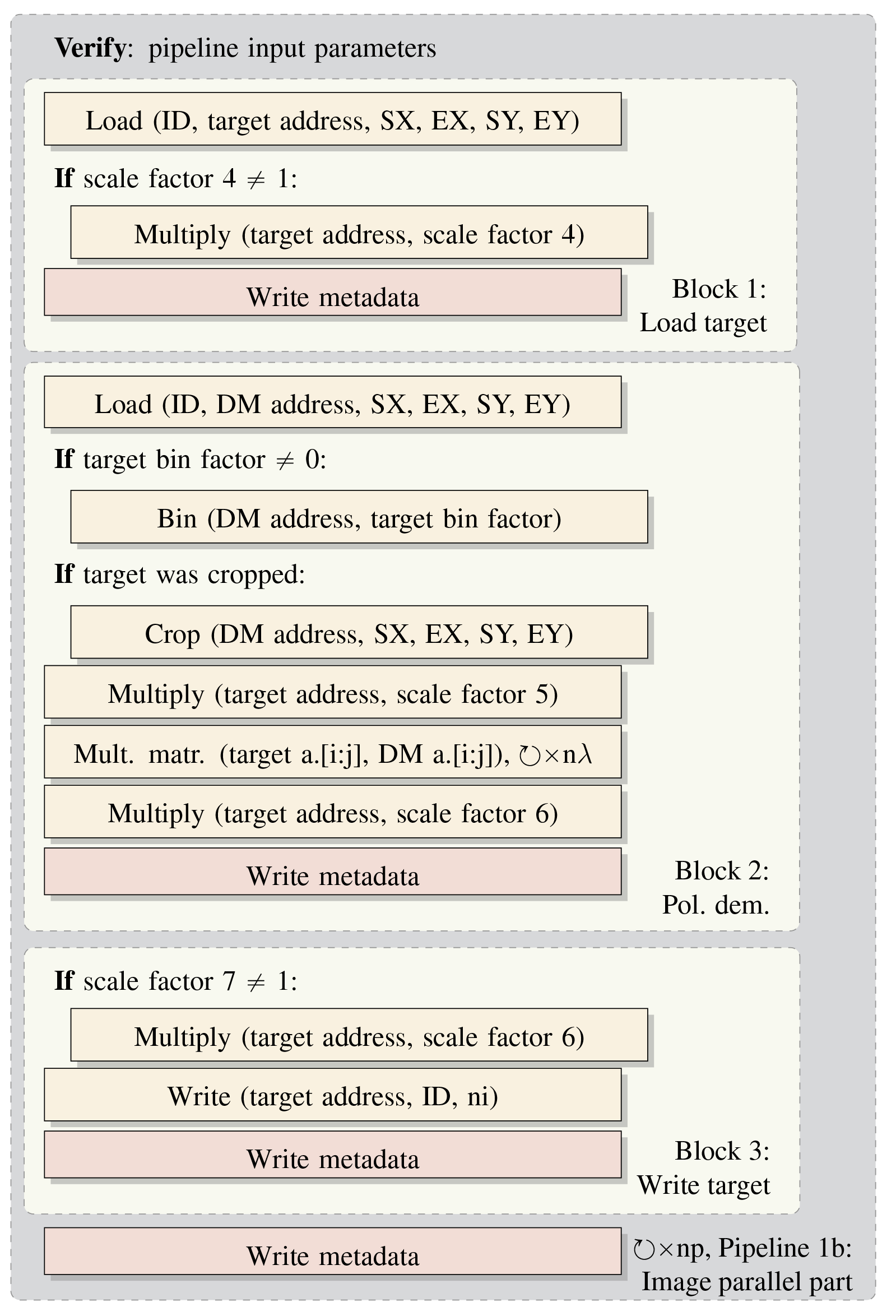}
		\caption{Image parallel part.}
		\label{fig:par}
	\end{subfigure}
	\caption[Pipelines.]%
	{Example science data processing pipeline. In (a): the parameter ni denotes the number of images processed at a given time, while nl is the number of times the image linear portion of the pipeline has to run to process the whole dataset. In (b): n$\lambda$ denotes the number of wavelengths processed at a given time, while np denotes the iterations of the image parallel part of the pipeline necessary for the full processing of the dataset.}
	\label{fig:example}
\end{figure}

To demonstrate how the system works, we take an example of a science data pre-processing pipeline, consisting of 3 steps: dark field subtraction, flat field division, and polarimetric demodulation. This is the minimum that is needed for pre-processing, and would be followed by RTE inversion for full on-board analysis.\\
Each step becomes a block, and the whole pipeline is split into the image linear and image parallel part. Each of these starts with the loading of the target dataset, and ends with the storage of the result (see \cref{fig:example}). Each block scales the datasets to adapt to the defined scaling interface between them, and to calculate the results at the required accuracy. The blocks also check the history of the target dataset and adapt the calibration data applied to them. Additional verifications generate warnings to determine possible problem-sources on ground. Each pipeline is inside a loop and is executed the number of times necessary to process the whole dataset, when this is split into sub-parts.\\
\begin{figure}[p]
	\centering
	\begin{subfigure}[b]{.23\textwidth}
		\includegraphics[width=\textwidth]{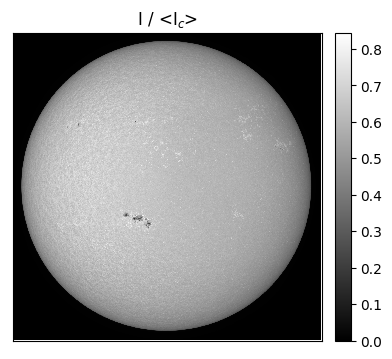}
		\caption{$I$}
		\label{fig:I}
	\end{subfigure}
	\begin{subfigure}[b]{.248\textwidth}
		\includegraphics[width=\textwidth]{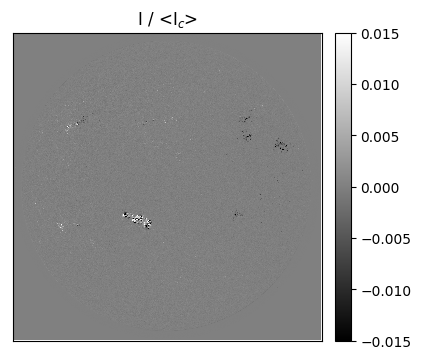}
		\caption{$Q$}
		\label{fig:Q}
	\end{subfigure}
	\begin{subfigure}[b]{.248\textwidth}
		\includegraphics[width=\textwidth]{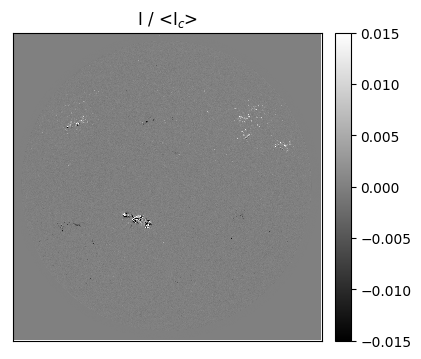}
		\caption{$U$}
		\label{fig:U}
	\end{subfigure}
	\begin{subfigure}[b]{.248\textwidth}
		\includegraphics[width=\textwidth]{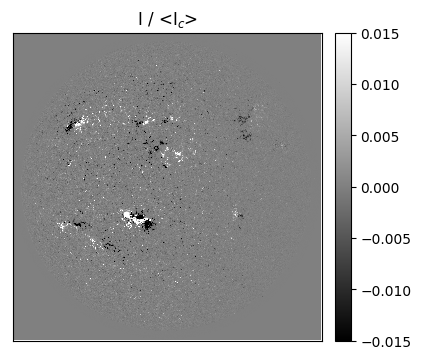}
		\caption{$V$}
		\label{fig:V}
	\end{subfigure}

	\begin{subfigure}[b]{.24\textwidth}
		\includegraphics[width=\textwidth]{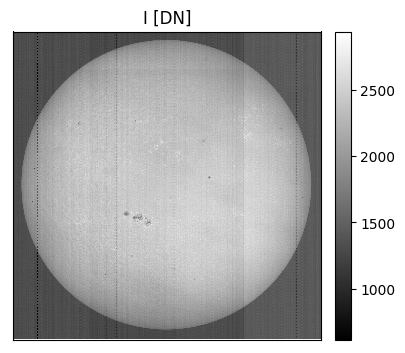}
		\caption{I$_1$}
		\label{fig:I1}
	\end{subfigure}
	\begin{subfigure}[b]{.24\textwidth}
		\includegraphics[width=\textwidth]{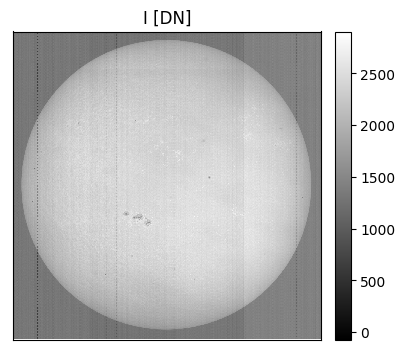}
		\caption{I$_2$}
		\label{fig:I2}
	\end{subfigure}
	\begin{subfigure}[b]{.24\textwidth}
		\includegraphics[width=\textwidth]{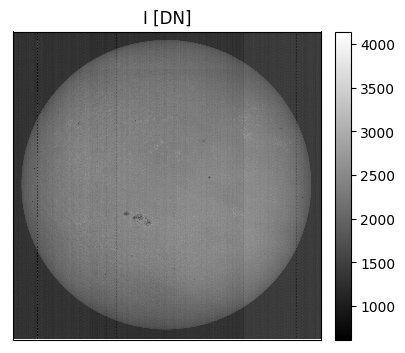}
		\caption{I$_3$}
		\label{fig:I3}
	\end{subfigure}
	\begin{subfigure}[b]{.24\textwidth}
		\includegraphics[width=\textwidth]{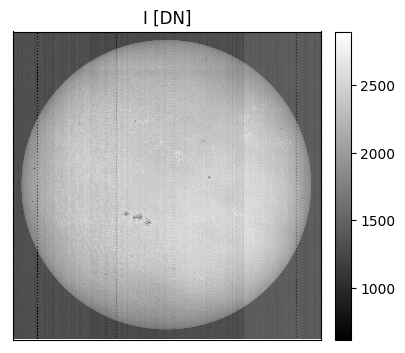}
		\caption{I$_4$}
		\label{fig:I4}
	\end{subfigure}
	\vspace{.2 cm}
	\caption{Test dataset at line centre. (a) - (d): Stokes images from SOPHISM\cite{blanco2018sophism}. (e) - (h): Input to the pipeline, Stokes images modified according to \cref{eq:inputdata}. The input images show the dark field (the stripes), and the flat field (smudges and dust grains). These effects are compensated by applying the calibration data shown in \cref{fig:Calib}, then demodulated to arrive at the Stokes images once again.}
	\label{fig:Input}
\end{figure}\begin{figure}[p]
	\centering
	\begin{subfigure}[b]{.246\textwidth}
		\includegraphics[width=\textwidth]{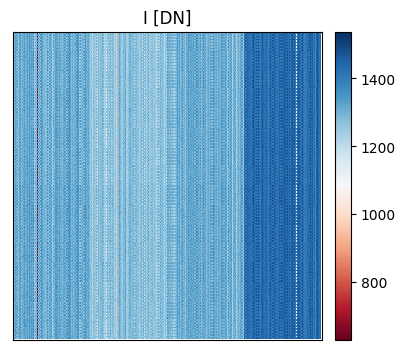}
		\caption{Dark field.}
		\label{fig:dark}
	\end{subfigure}
	\begin{subfigure}[b]{.229\textwidth}
		\includegraphics[width=\textwidth]{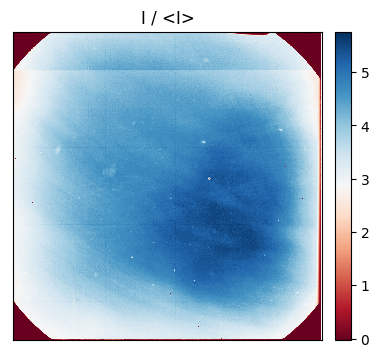}
		\caption{Flat field.}
		\label{fig:flat}
	\end{subfigure}\hspace{5pt}
	
	\begin{subfigure}[b]{.3\textwidth}
		\includegraphics[width=\textwidth]{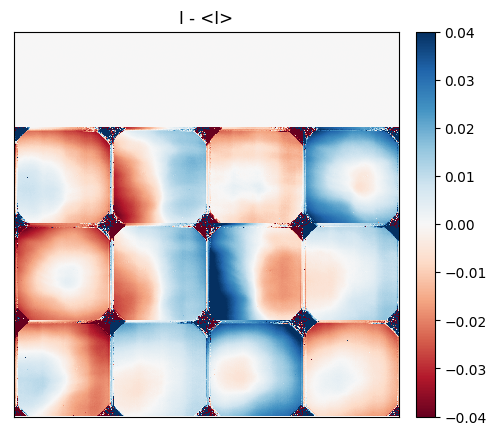}
		\caption{DM Uniformity across FOV.}
		\label{fig:MM}
	\end{subfigure}
	\vspace{.2 cm}
	\caption{The calibration data used in the test run is recorded in the laboratory. (a): The dark field shows the characteristic sensor pattern: four distinct vertical stripes and row-to-row horizontal variations. (b): The flat field used in the tests is the same for all wavelengths and modulation states. (c): The variation of the 4$\times$4 demodulation matrix elements across the FOV, treated as a dataset of 16 images.}
	\label{fig:Calib}
\end{figure}\begin{figure}[p]
	\centering
	\begin{subfigure}[b]{.24\textwidth}
		\includegraphics[width=\textwidth]{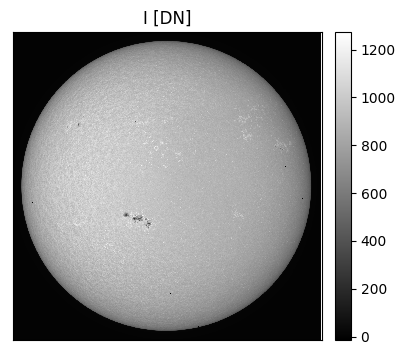}
		\caption{I$_1$}
		\label{fig:I1r}
	\end{subfigure}
	\begin{subfigure}[b]{.245\textwidth}
		\includegraphics[width=\textwidth]{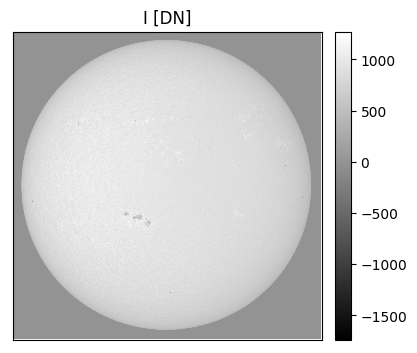}
		\caption{I$_2$}
		\label{fig:I2r}
	\end{subfigure}
	\begin{subfigure}[b]{.239\textwidth}
		\includegraphics[width=\textwidth]{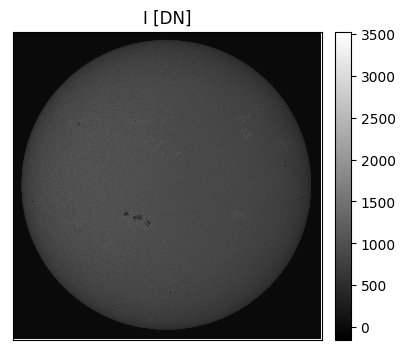}
		\caption{I$_3$}
		\label{fig:I3r}
	\end{subfigure}
	\begin{subfigure}[b]{.239\textwidth}
		\includegraphics[width=\textwidth]{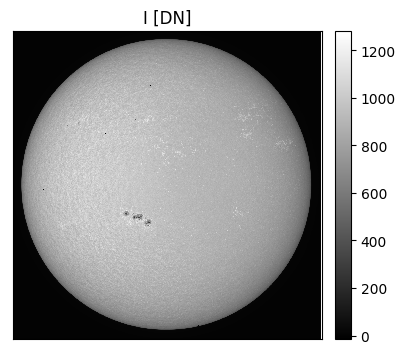}
		\caption{I$_4$}
		\label{fig:I4r}
	\end{subfigure}

	\begin{subfigure}[b]{.24\textwidth}
		\includegraphics[width=\textwidth]{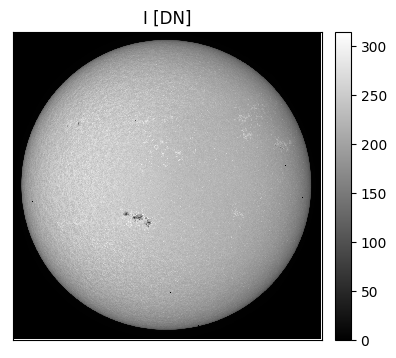}
		\caption{$I$}
		\label{fig:Ir}
	\end{subfigure}
	\begin{subfigure}[b]{.24\textwidth}
		\includegraphics[width=\textwidth]{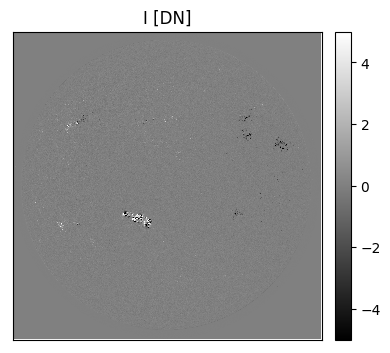}
		\caption{$Q$}
		\label{fig:Qr}
	\end{subfigure}
	\begin{subfigure}[b]{.24\textwidth}
		\includegraphics[width=\textwidth]{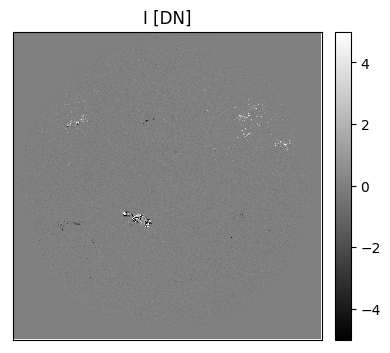}
		\caption{$U$}
		\label{fig:Ur}
	\end{subfigure}
	\begin{subfigure}[b]{.24\textwidth}
		\includegraphics[width=\textwidth]{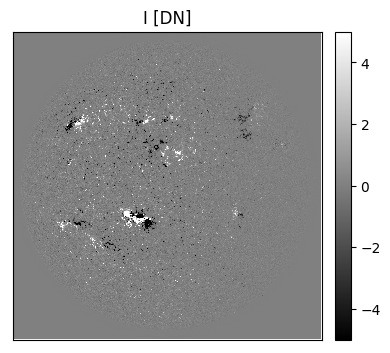}
		\caption{$V$}
		\label{fig:Vr}
	\end{subfigure}
	
	\begin{subfigure}[b]{.242\textwidth}
		\includegraphics[width=\textwidth]{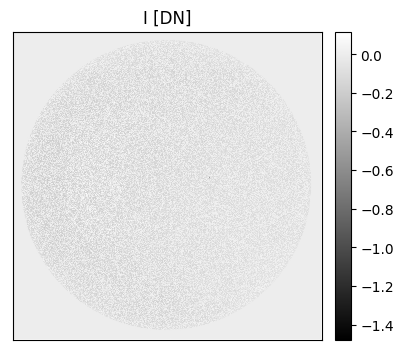}
		\caption{$\delta I$}
		\label{fig:EIr}
	\end{subfigure}
	\begin{subfigure}[b]{.244\textwidth}
		\includegraphics[width=\textwidth]{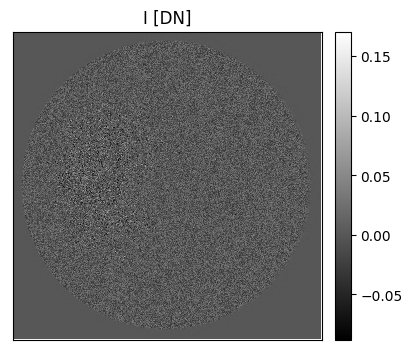}
		\caption{$\delta Q$}
		\label{fig:EQr}
	\end{subfigure}
	\begin{subfigure}[b]{.244\textwidth}
		\includegraphics[width=\textwidth]{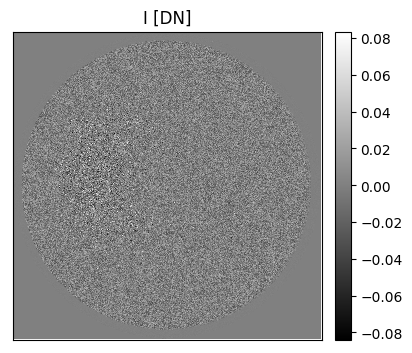}
		\caption{$\delta U$}
		\label{fig:EUr}
	\end{subfigure}
	\begin{subfigure}[b]{.244\textwidth}
		\includegraphics[width=\textwidth]{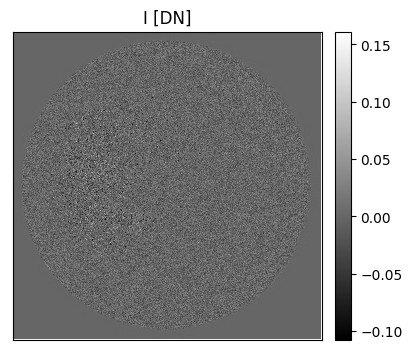}
		\caption{$\delta V$}
		\label{fig:EVr}
	\end{subfigure}

	\vspace{.2 cm}
	\caption{Result dataset at line centre. (a) - (d): Input images corrected for the effects of dark and flat field. (e) - (h): Result of the demodulation of the images in \cref{fig:I1r} to \ref{fig:I4r}. (i) - (l): The difference after demodulation when executing the pipeline in floating point representation on a desktop, and executing it in fixed point representation on SO/PHI. The differences occur inside the solar disk in form of quantisation noise. (The units are in fixed point DN-s.)}
	\label{fig:Output}
\end{figure}The input data for the demonstration is generated from images taken by the SDO/HMI instrument\cite{Schou2012HMI}, transformed into Stokes images with the SO/PHI instrument simulator, SOPHISM\cite{blanco2018sophism}. See \cref{fig:I} to \ref{fig:V}. \newpage \noindent These images are further manipulated to represent the raw observables, according to the following equation:
\begin{equation}
 \bm I^\text{input}_{\lambda} \unboldmath(x,y)= c_1 c_2 [\bm M(x,y) \cdot \bm I^\text{Stokes}_{\lambda}(x,y)]   I^\text{Flat}(x,y) + I^\text{Dark}(x,y),
\label{eq:inputdata}\end{equation}
where "$\cdot$" stands for matrix multiplication, $\lambda$ denotes the wavelength, $x$ and $y$ denote the image row and column size, $\bm I^\text{input}_\lambda$ is the created test dataset, a $1\times4$ matrix of images at $6$ different wavelengths, $\bm I^\text{Stokes}_\lambda$ is the $1\times4$ matrix of Stokes images created with SOPHISM at 6 wavelengths, which can be expressed for any of the wavelengths as: \[\bm I^\text{Stokes} = [I, Q, U, V]^T,\] $\bm{M}$ is the instrument Modulation Matrix, a $4\times4$ matrix of images, describing how the instrument transforms the sought Stokes images into practically measurable light levels for each pixel of its field of view (FOV), applied for all wavelengths, $c_1$ is the constant that scales the normalised images to represent the number of incident photons in 20\,ms exposure time, $c_2$ is the constant that converts the pixels from number of photons accumulated on the detector to Digital Numbers (DN-s), ${I}_\text{Flat}$ is one image, the normalised flat field of the telescope, applied to all wavelengths and polarisation states, and ${I}_\text{Dark}$ is the dark field image of the sensor at 20\,ms integration time in DN-s, the same for all wavelengths and polarisation states.\\
\begin{table}[t]
	\small
	\caption{The metadata from the image linear pipeline portion shows all executed blocks, ending with the entry by the pipeline. Parameter 1 shows the ID of the Target Dataset (TD), Dark Field (DF), Flat Field (FF) and Result Dataset (RD). The dataset is processed in the RFPGA, treating all 24 images (see image start (S) and end (E)), and the full field of view (see row and column start (S) and end (E)) at once. The flat correction block generates a warning about produced NaN-s, which is also shown in the pipeline return.} 
	\label{tab:metadata}
	\begin{center}       
		\begin{tabular}{|l|l|l|l|l|l|l|l|l|l|} 
			\hline
			Return value & Function ID & param. 1 & param. 2 & image S & image E & row S & row E & col. S & col. E \\
			\hline
			Correct & Load & TD ID & 0 & 0 & 23 & 0 & 2047 & 0 & 2047\\
			\hline
			Correct & Correct Dark  & DF ID & 0 & 0 & 23 & 0 & 2047 & 0 & 2047\\
			\hline
			W: NaNs & Correct Flat  &  FF ID & 0 & 0 & 23 & 0 & 2047 & 0 & 2047\\
			\hline
			Correct & Store  & RD ID & 0 & 0 & 23 & 0 & 2047 & 0 & 2047\\
			\hline			
			W: NaNs & Linear ppln. & RD ID & 0 & 0 & 23 & 0 & 2047 & 0 & 2047\\
			\hline
		\end{tabular}
	\end{center}
\end{table}The pipeline removes the dark field and the flat field from the input images, then demodulates them to arrive to the Stokes images. The calibration data applied by the pipeline is recorded in the laboratory, using the SO/PHI Flight Model (see \cref{fig:Calib}). The dark field is the offset of the image detector, while the flat field shows the gain variations across the FOV caused by the optics. The Demodulation Matrix is the inverse of the Modulation Matrix used in \cref{eq:inputdata}. The test is run on a model of SO/PHI, which is representative in its data processing unit.\\
We compare the output of the pipeline run on SO/PHI to results produced on the desktop with floating point calculations on the same input data. We expect these to give similar results with differences within an error margin, coming from the pixel quantisation in fixed point notation. The differences are as we expect them, appearing as quantisation noise over the solar disk. The standard deviation of the differences is 0.05\% for the Stokes $I/I_r$ image, and 6\%, 5.3\% and 1.8\% for $Q/Q_r$, $U/U_r$ and $V/V_r$, respectively, where $I_r$, $Q_r$, $U_r$ and $V_r$ denote the resulting Stokes images from the reference pipeline. The polarimetric errors coming from the fixed point representation are 0.017\%, 0.015\% and 0.016\% for $Q / I_{c}$, $U / I_{c}$ and $V / I_{c}$, respectively, where $I_{c}$ is the Stokes $I$ at continuum wavelength. These values are within the SO/PHI scientific requirements, and may be further optimised by adjusting the scaling parameters.\\
The metadata entries, recorded to summarise the blocks and pipelines executed on the dataset, contain their ID, given parameters and return values. See \cref{tab:metadata}. To keep a complete record, in addition to these, entries from other sources are also logged (e.g. entries from the image processing operations, or describing the dataset).
\newpage
\section{Conclusions}
SO/PHI is the first solar imaging spectropolarimeter to perform autonomous on-board instrument calibration and autonomous data reduction. The data processing involved with these tasks is implemented in a software framework distributed between a microprocessor and two RFPGAs. The microprocessor controls the RFPGAs, which implement functions with long computational times to accelerate them (e.g. image processing functions). The framework is designed to ensure the implementation of a large variety of algorithms, and their easy modification when necessary, while operating with limited memory resources. To ensure that the autonomous process can be fully recovered on ground, a metadata log is kept about all operations performed, their parameters and return values. A warning system is implemented for supporting the scientists on ground to find possible errors in the processing. The presented results demonstrate the integrity of the system and its ability to process the data.\\
The data processing software frame has the following key features:
\begin{itemize}
	\item block approach,
	\item pipeline implementation of tasks,
	\item parametrised pipelines for processing data in sub-sets,
	\item wide error detection and handling,
	\item a system for maintaining required accuracy during fixed point operations,
	\item extensive metadata logging.
\end{itemize}
The software system was developed for the specific needs of the SO/PHI instrument on-board the Solar Orbiter, to cope with the telemetry limitations. However, due to the adaptability of the system, it is planned to serve as basis for on-board data processing systems for future instruments in similarly challenging orbits. A more detailed description of the SO/PHI data processing system will be given in a paper currently in preparation.

\section*{Acknowledgements}
This work was carried out in the framework of the International Max Planck Research School (IMPRS) for Solar System Science at the Max Planck Institute for Solar System Research (MPS). Solar Orbiter is a mission lead by the European Space Agency (ESA) with significant contribution from National Aeronautics and Space Administration (NASA). The SO/PHI instrument is supported by the German Aerospace Center (DLR) through Grant 50 OT 1201. The Spanish contribution has been partly funded by the Spanish Research Agency under project ESP2016-77548-C5, partially including European FEDER funds. The solar data used in the test are the courtesy of NASA/SDO HMI science team.

\bibliography{report.bib} 
\bibliographystyle{spiebib.bst} 

\end{document}